# Fabrication of Embedded Microvalve on PMMA Microfluidic Devices through Surface Functionalization

A. G. G. Toh, Z.F. Wang and S.H. Ng

Singapore Institute of Manufacturing Technology

*Abstract-* The integration of a PDMS membrane within orthogonally placed PMMA microfluidic channels enables the pneumatic actuation of valves within bonded PMMA-PDMS-PMMA multilayer devices. Here, surface functionalization of PMMA substrates via acid catalyzed hydrolysis and air plasma corona treatment were investigated as possible techniques to permanently bond PMMA microfluidic channels to PDMS surfaces. FTIR and water contact angle analysis of functionalized PMMA substrates showed that air plasma corona treatment was most effective in inducing PMMA hydrophilicity. Subsequent fluidic tests showed that air plasma modified and bonded PMMA multilayer devices could withstand fluid pressure at an operational flow rate of 9μL/min. The pneumatic actuation of the embedded PDMS membrane was observed through optical microscopy and an electrical resistance based technique. PDMS membrane actuation occurred at pneumatic pressures of as low as 10kPa and complete valving occurred at 14kPa for ~100μm × 100μm channel cross-sections.

## I. INTRODUCTION

Microfluidic devices offer an attractive route towards the miniaturization of chemical and biological processes in lab-on-chip applications. In these devices, control of micro- to nano-litre fluid volumes in processes such as fluid transport, separation, mixing, and reaction can be achieved through mechanical valving techniques. Traditionally, 3D valve structures are fabricated on silicon-based devices through bulk or surface micromachining methods [1]. Recently, multilayer "soft lithography" methods have been developed to fabricate the microfluidic channels and pneumatic microvalves through replication molding of soft elastomeric materials [2]. The valving is realized through expending the elastomeric membrane into microfluidic channel, which gradually pinches the flow under certain applied air pressures. Polydimethylsiloxane (PDMS) is the popular material choice for replication as it can be easily molded and has a Young's modulus of ~750 kPa that allows for efficient valving at low actuation pressures [3]. However, such PDMS based multilayer valves suffer from an important drawback: the swelling of PDMS channels after prolonged fluid exposure [4]. The swelling of channels subsequently results in inter-layer misalignment that affects the proper functioning of the entire microfluidic device, especially in microarray devices. In addition, the solvent induced swelling of the material limits the application range of such PDMS devices due to the concern of contamination.

As such, rigid thermoplastic polymers such as poly (methyl methacrylate) (PMMA) have been investigated as a possible substrate for the fabrication of microfluidic devices due to their low susceptibility to swelling and biocompatibility [5]. Additionally, open microchannels within PMMA substrates can be rapidly prototyped using high throughput methods such as hot or cold embossing, laser ablation or micro-injection molding. In order to build multilayer embedded valves for lab-on-chip applications, the fabricated PMMA microfluidic substrates must subsequently be bonded to sandwich an elastomeric membrane.

Here, we developed a pneumatically active embedded valve capable of actuation at low pressures within a PMMA multi-stacked microfluidic configuration. The principle of valve actuation is based on pressurizing/depressurizing an air chamber that supports a moving elastomeric membrane located below a PMMA fluidic channel (see Figure 1). The valve configuration is designed to accommodate valve arrays within a PMMA-PDMS-PMMA bonded microfluidic network. Low temperature surface functionalization methods such as air plasma and acid hydrolysis treatment were investigated to improve the bonding between the PMMA microfluidic substrates and PDMS based elastomeric thin membrane. In order to examine the valving efficiency, we developed a method for detecting valve efficiency (or channel constriction) based on electrical resistance changes of an electrolyte flowing within microfluidic channels.

## II. EXPERIMENTAL

*Materials and chemicals*
Poly(methylmethacrylate) (PMMA) sheets of thickness 1.5mm were purchased from Dama Enterprise (Singapore) and cut to circular discs of diameter 101.6mm (~4") before hot embossing of microfluidic channels. SU-8 3025 negative photoresist were purchased from Microchem and used as received. PDMS membranes were fabricated from Slygard 184 silicone elastomer and cure (both purchased from Dow Corning). Sulphuric acid (ACS reagent grade) was obtained from Merck. Electrolytic fluid solution was obtained by diluting 2ml of food dye in 15ml of deionized water (with conductivity of 0.55mS) and subsequently adding 0.5g of Potassium chloride (KCl) salt (purchased from Metrohm).

*Fabrication of PMMA microfluidic chips*

PMMA microfluidic chips were fabricated via hot embossing using a platen hot embossing machine. The hot embossing machine used was a bench-top hydraulic press with a maximum loading of 15 tons. The hydraulic press compressed the embossing tool and substrate between two heated platens. During embossing, the embossing tool and substrate are aligned between the top and bottom platens at room temperature. The platens were subsequently heated above the glass transition temperature ($T_g$) of PMMA. Once the platens were heated to the required temperatures, a pressure of 6MPa was applied for 20mins. The platens were then cooled to 50ºC and the embossed chips were retrieved. The embossed chips were then diced to rectangular pieces of 75mm × 55mm.

Prior to hot embossing, the Si/SU-8 embossing tool was fabricated via standard photolithography processes. SU-8 photoresist was spin coated at 1000rpm to yield a thickness of ~100μm. Photolithography of channel patterns on the SU-8 layer was performed using the Karl Suss mask aligner.

Pneumatic control channels of a trapezoid cross section were micro-machined using a LPKF circuit board plotter. The base channel width and total channel depth was measured to be ~100 μm and ~250 μm respectively. Markers on both the PMMA embossed chips (fluidic channels) and the machined pneumatic control channels were used to align the substrates during bonding. The entire microfluidic chip fabrication process is summarized in Fig. 1.

*Fabrication of PDMS membrane valves*

The PDMS elastomer and curing agent were mixed in a 3:1 ratio to yield a PDMS prepolymer solution. The prepolymer was degassed *in vacuum* for 1 hour before being spin coated on the adhesive layer (see Fig. 1); the adhesive layer had previously been tape rolled onto the PMMA substrate consisting of pneumatic control channels. The PDMS membrane was spun at 1000rpm for 60secs and subsequently degassed again for 1 hour. The PDMS membrane was then completely cured at 65ºC for 3 hours.

*Surface functionalization of PMMA and PDMS surfaces*

Two types of PMMA surface functionalization processes were investigated: acid catalyzed hydrolysis and air corona treatment. During acid catalyzed hydrolysis, PMMA embossed chips were sonicated in 1M of sulphuric acid at 60ºC for 20mins. After removal from the sulphuric acid bath, the PMMA chip was cleaned in deionized water. Plasma treatment of the PMMA chips was performed in air using a corona plasma machine. The plasma treatment was performed for 120 secs at an applied power of 0.5 kW and a frequency of 10 kHz.

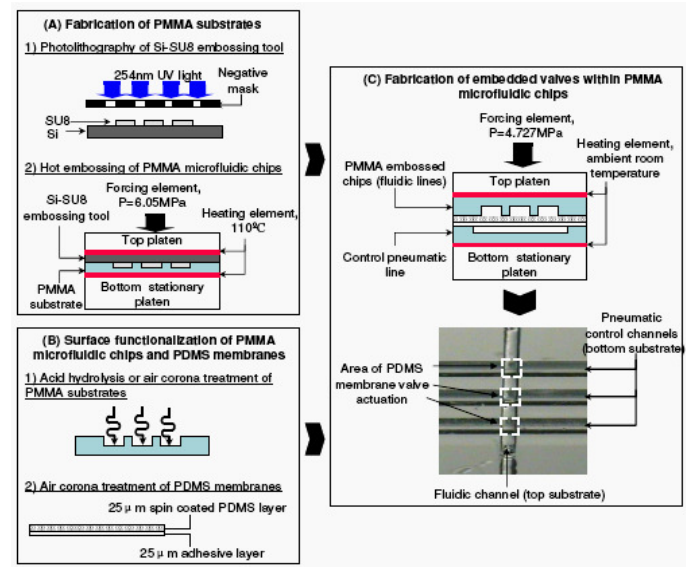

Fig. 1 Processes involved in fabricating embedded PDMS valves within PMMA microfluidic chips.

*Characterization of PDMS-PMMA microfluidic chips*

Dimensional fidelity of Si/SU-8 embossing tools and PMMA embossed chips were measured using a Talyscan mechanical profilometer. PMMA surface functionalization was examined through water contact angle measurements using the VCA Optima XE, which had a measurement accuracy of ±1°. Deionized water was used as the probe liquid in all measurements. The advancing water contact angles presented are an average of 3 measurements.

The change in PMMA surface chemistry after surface functionalization was observed using a Bio-Rad Excalibur Fourier transform infrared (FTIR) spectroscope. The FTIR spectra were collected from 64 measurement scans at a resolution of 4cm$^{-1}$ within the wavenumber range of 1000-1800cm$^{-1}$.

### III. RESULTS AND DISCUSSION

*Hot embossing of PMMA substrates*

In order to obtain good fidelity between the fabricated Si/SU-8 embossing tool and PMMA substrates, hot embossing of PMMA microfluidic chips was attempted at a temperature range of 95 - 120ºC. This temperature range was investigated based on previously reported $T_g$ (~105ºC) values of bulk PMMA substrates.[6, 7] Effective embossing was found to occur at a temperature of 110ºC, an applied pressure of 6 kN/m$^2$ (~5 tons of applied mass) and a duration of 20mins. At the aforementioned parameters, profilometric measurements showed that the embossed channels had a dimensional fidelity to the Si/SU-8 embossing tool of greater than 90% (see Fig. 2, 3 and Table 1). While we found that a higher embossing temperature of 120ºC produced PMMA chips with better dimensional fidelity to the embossing tool, a slight warpage of the PMMA substrates was observed upon cooling of the platens to 50ºC. An embossing temperature of 110ºC was therefore found to be an effective compromise between dimensional fidelity and substrate flatness.

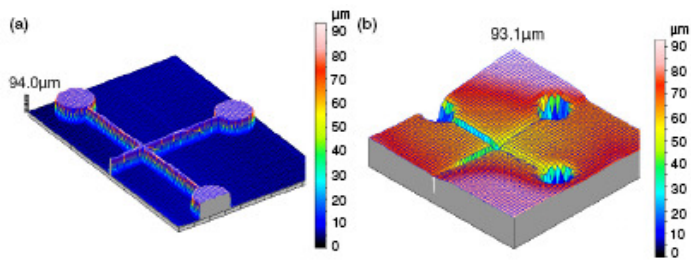

Fig. 2 3D mechanical stylus measurement area of 10mm × 8mm of the (a) Si/SU-8 embossing mold and (b) embossed PMMA microfluidic chip at optimized embossing parameters.

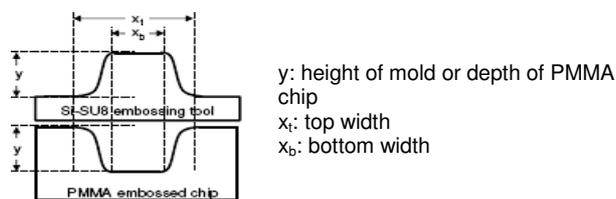

Fig. 3 Channels dimensions measured by mechanical stylus measurements listed in Table 1.

TABLE I
DIMENSION COMPARISONS BETWEEN THE SI/SU-8 EMBOSSING TOOL AND PMMA EMBOSSED CHIPS (SAMPLE SIZE: n=3)

| Sample | Temperature (ºC) | $x_t$ (μm) | $x_b$ (μm) | y (μm) |
|---|---|---|---|---|
| Si/SU-8 embossing tool | - | 98.2 ± 0.9 | 94.8 ± 2.1 | 94.0 ± 1.2 |
| PMMA embossed chips | 95 | 78.4 ± 3.1 | 57.7 ± 4.6 | 68.4 ± 4.3 |
|  | 100 | 82 ± 2.7 | 69.4 ± 2.9 | 72.3 ± 3.8 |
|  | 110 | 93.6 ± 2.0 | 85.0 ± 1.9 | 87.1 ± 1.9 |
|  | 120 | 96.7 ± 1.6 | 86.5 ± 2.4 | 91.0 ± 2.4 |

*Surface functionalization of embossed PMMA chips*

The surface of the embossed PMMA chips was functionalized through two methods: acid catalyzed hydrolysis and air corona plasma treatment. Sulfuric acid catalyzed hydrolysis is considered to be a random process whereby carboxylate or ester terminal groups are catalyzed via the nucleophilic attack of hydroxylic groups. These carboxylate or ester terminal groups are known to subsequently form carboxylic acid groups that render the PMMA surface more hydrophilic than its native state [8]. On the other hand, air plasma treatment of PMMA surfaces is thought to occur due to the radical attack and ozonation of PMMA surface groups [9]. As the gas molecule bombardment of PMMA surfaces is likely to produce a variety of modified surface groups, the process of PMMA hydrophilization via air plasma treatment is less understood. The hydrophilization effects of these treatments could be observed via water contact angle measurements of the modified PMMA surfaces as shown in Fig. 4. The average advancing water contact angle of native unmodified PMMA was observed to 79.2 ± 5.3º. This value is in agreement to those for unmodified PMMA bulk surfaces reported previously.[8] Acid catalyzed hydrolysis and plasma treatment of the PMMA surface was found to reduce the average advancing water contact angle to 69.8 ± 3.8º and 37.2 ± 2.7º respectively. The surface chemistry modification on the observed PMMA hydrophilization was further analyzed via FTIR spectroscopy.

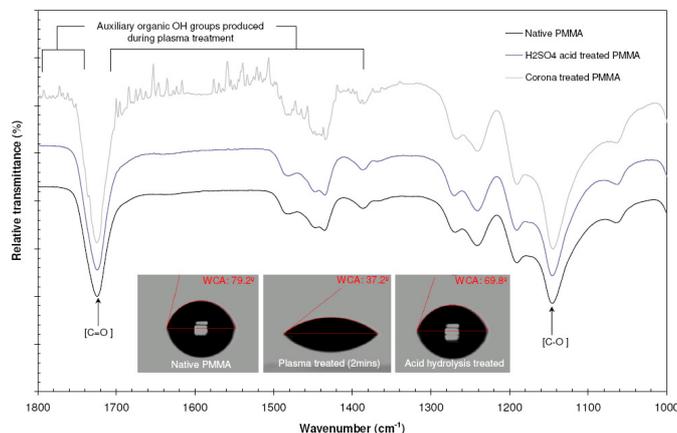

Fig. 4 Surface functionalization of PMMA substrates to aid bonding between the PDMS membrane valve layer to PMMA microfluidic layers. Inset shows the change in advancing water contact angles before and after air plasma and $H_2SO_4$ acid hydrolysis treatment of PMMA surfaces.

FTIR spectroscopic analysis of the treated surfaces indicated that air plasma treatment resulted in larger PMMA surface modifications, hence explaining the greater surface hydrophilization observed (see Fig. 4). It can be seen from Fig. 4 that the C-O peak areas at wavenumber of 1148cm$^{-1}$ and 1730 cm$^{-1}$ [5] increased significantly after plasma treatment as compared to acid catalyzed hydrolysis treatment. The increase in C-O surface groups results in an increase in surface electronegativity and polarity, hence increasing its affinity to water. Interestingly, an increase in the number of absorption bands from wavenumber 1350cm$^{-1}$ to 1800cm$^{-1}$ was also observed for PMMA substrates that had been plasma treated for 120 secs. This suggests the presence of multiple auxiliary organic groups on the modified PMMA surface, each likely contributing to the surface polarity [8, 9]. It should be noted that an increase in plasma treatment time of up to 300 secs did not result in significant increases in hydrophilization and caused slight warpage and subsequent cracking of the embossed PMMA chips during compression assembly. The increase in PMMA substrate hardness during oxygen air plasma has previously been observed and quantified through nano-indentation experiments [8]. It is likely that air plasma treatment results in the formation of similar oxide and carboxylic species being formed during oxygen plasma treatment. These species have been attributed to the formation of a brittle surface layer on oxygen treated PMMA bulk substrates [8]. Nonetheless, at an air plasma treatment duration of 120 secs, the incorporation of oxides (C-O) and auxiliary aldehydes/organic OH groups to the PMMA surface is expected to improve surface polarity and hence aid bonding between the PDMS membrane and PMMA substrate.

*Room temperature bonding of PMMA substrates*

Bonding between surface modified PMMA embossed chips, PDMS membranes and PMMA control channel substrates were performed by room temperature compression (see Fig. 1). Effective bonding of the PDMS membrane to plasma treated PMMA embossed chips was achieved at an applied tonnage range of 1.5-2.0 ton (this corresponds to an applied pressure range of 3.57-4.23 MPa for a chip area of 41.25 cm$^2$). At this pressure range, fluid and air flow within the fluid and pneumatic control channels respectively did not exhibit clogging. Furthermore, fluid dye was able to fill the fluidic channels at a flow rate of up to 9 μL/min without any observable leakage.

On the contrary, PMMA embossed chips that were functionalized via the acid catalyzed hydrolysis route did not bond well to the PDMS membrane regardless of the applied compression pressure. Additionally, at larger compression pressures of ~7 MPa and above, no effective bonding was achieved and the pressure was found to 'squeeze' the membrane into the pneumatic control and fluidic channels as shown in Fig. 5. This suggests that bonding was no longer about physical contact between the PDMS and PMMA surfaces, but physio-chemical interactions between the two surfaces. Here, air plasma corona treatment of the PMMA chip was found to result in better bonding with the PDMS membrane as compared to an acid catalyzed hydrolysis treated chip. We also observed that bonding was not possible without prior air plasma corona treatment.

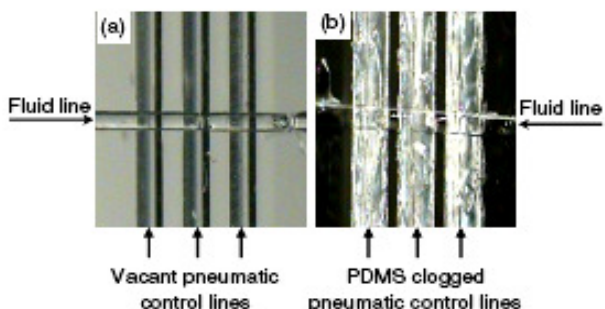

Fig. 5 Micrograph of room temperature compression between PDMS membranes and PMMA substrates at (a) 4.23 MPa (for air plasma corona treated PMMA) and at (b) 7.13 MPa (for acid catalyzed hydrolysis treated PMMA).

*PDMS microvalve actuation*

Once effective bonding between the PDMS membrane layer and PMMA substrates was achieved, the chip was assembled and fitted with inlet/outlet connectors (supplied by Nanoport). Electrical wiring within the inlet/outlet connectors were fitted for electrical resistance sensing of valving efficiency. (see Fig 6(a)) During fluidic tests, top fluidic channels were connected to syringe pumps while bottom pneumatic control channels were connected to compressed air supply (see Fig. 6 (a, b) for microfluidic chip assembly). The "push up" valving mechanism occurring during pressurizing of the pneumatic air control channel is schematically shown in Fig. 6 (c).

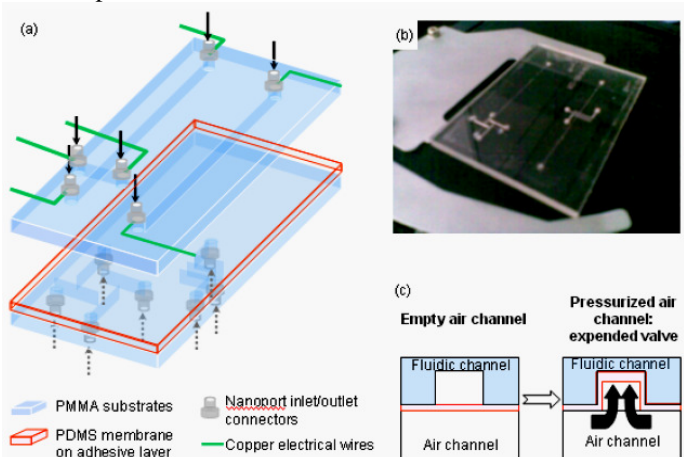

Fig. 6 (a) Schematic assembly of top PMMA embossed microfluidic chips and, bottom PDMS membrane layer and orthogonally placed PMMA pneumatic channels. (solid arrows indicate fluid inlet/outlets while dotted arrows indicate air inlet/outlets). (b) photo of the final assembled PMMA-PDMS-PMMA microfluidic chip and (c) illustration of PDMS membrane "push up" valve actuation during pressurizing of bottom air channels

The pneumatic actuation of the PDMS embedded membrane valve was observed through optical microscopy and is shown in Fig. 7. An increase in pneumatic control pressures to 34 kPa resulted in large deflections of the membrane valve, hence resulting in the constriction of fluid flow through the PMMA channel. Normal fluid flow could be resumed when the air chamber pressure was released. The device was able to withstand pressures resulting from the complete valving of fluid flow rates of 1 to 5μL/min, while the complete valving of flows rates greater than 5 μL/min, resulted in flow leakage. Further characterization of these membranes would provide more information on the pressure limitations of both the membrane and device.

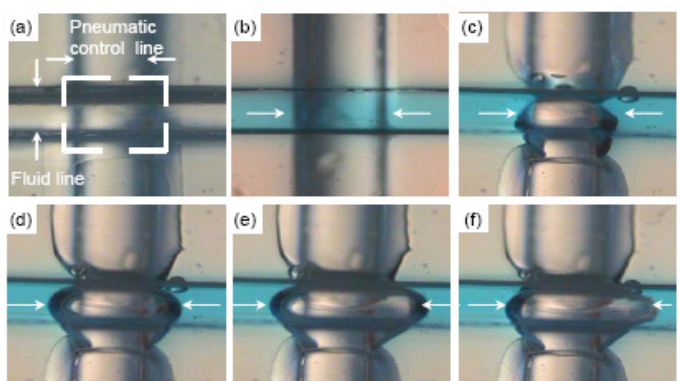

Fig. 7 Micrograph showing pneumatic control of blue electrolytic fluid dye (at a flow rate of 1μL/min) via a single embedded PDMS based microvalve during (a) no fluid flow, (b)open valving, and valving actuated at (c) P=10kPa, (d) P=14kPa, (e) P=29kPa, and (f) P=34kPa. The top horizontal channel is the PMMA fluid channel while the bottom vertical channel is the pneumatic control line. White arrows in (b) – (f) indicate the location of membrane actuation.

While optical visualization of the PDMS membrane movement provided valuable information on valving efficiency, the method proved to be restrictive for field testing of the device. Here, we developed a method to sense valving based on electrical resistance changes in the fluidic channel. Fig. 8 (a) and (b) illustrates the principle behind the electrical resistance measurement technique.

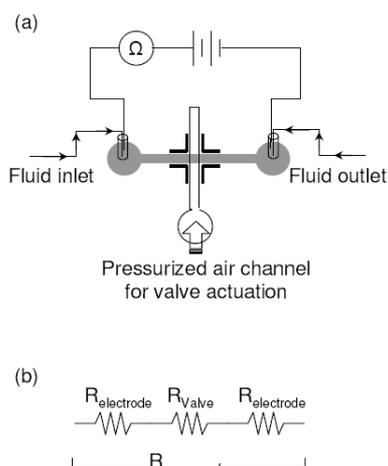

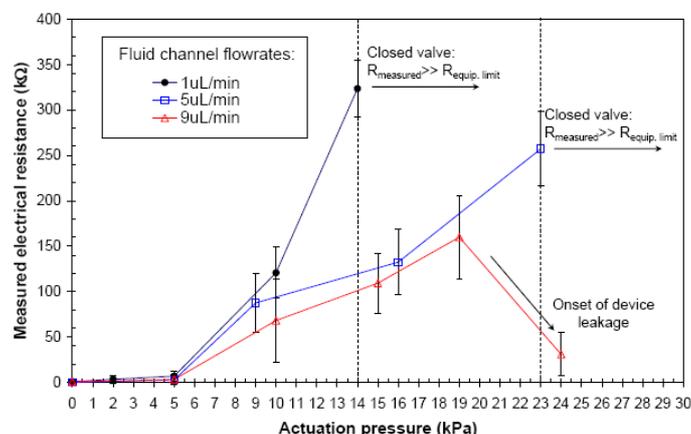

Fig. 8 (a) Electrical connections for a fluidic channel with a pneumatic channel located above its cross-section and (b) the electrical resistance model used to detect valving when an electrolyte flows through the fluidic channel.

Fig. 9 Change in electrical resistance as a function of PDMS membrane valve actuation for various flow rates of electrolytic fluid.

When a conductive fluid (electrolyte) is passed through the fluidic channel, a small measurement current can be applied across the channel length to induce a feedback voltage and hence electrical resistance. The fluid volume change in fluidic channel cross sectional area during valving is inversely related to the detected channel resistance (i.e. a decrease in channel cross sectional area results in an increase in channel resistance and vice versa). Fig. 9 shows the increase in resistance during valve actuation at various fluid flow rates. Presently, comparisons with optical micrographs (Fig. 7) show that this detection method can effectively measure complete open and closed valving. As observed from Fig. 9, the electrical resistance increases exponentially as the fluid valve closes; when the PDMS valve was completely closed, the electrical resistance detected increased to a value larger than 200MΩ, which was beyond the detection limit of our ammeter. Although the results show that an increase in valve actuation pressure (and hence channel restriction) results in an increase in detected electrical resistance, the non-linearity of the data suggests that the functional dependence between the two parameters are more complicated. Further experimental work and numerical calculations are needed before the valving states can be accurately mapped from electrical based resistance measurements and vice versa. Similarly, use of an AC power supply is likely to reduce electrical noise that was experienced during the electrical resistance detection under a DC power supply.

IV. CONCLUSIONS

PMMA microfluidic channels with high fidelity to Si/SU-8 embossing tools were achieved by optimizing the applied embossing temperature within 95 - 120 ºC. The subsequent fabrication of PDMS embedded valves within PMMA substrates was made possible through the oxidation of PMMA surfaces via air plasma corona treatment. Air plasma corona treatment for 120 secs decreased the advancing water contact angles of PMMA surfaces from 79.2 ± 5.3º to 37.2 ± 2.7º. The hydrophilic PMMA substrates were considered to result in an improvement in bonding with the PDMS membrane layer. Subsequent fluid test of the multi-stacked PMMA/PDMS/PMMA device proved that the bond strength achieved was able to withstand fluid flow rates of up to 9 µL/min.

The multi-stacked PMMA/PDMS/PMMA device was used to demonstrate the expansion/retraction of PDMS valves within fluidic channels. Actuation of the PDMS valves was achieved at air channel pressures of as low as 10 kPa and complete fluidic channel sealing could be achieved at air pressures of between 14 to 23 kPa. A novel electrical resistance-based detection of valve efficiency successfully monitored the complete open and closed valving states in real-time. The presented solution of such embedded valves has increasing relevance as the need for fluid valving within disposable microfluidic chips (to prevent sample cross contamination) has been well identified for many lab-on-a-chip processes.

ACKNOWLEDGMENT

This research is funded by the Agency for Science, Technology and Research (A*STAR), Singapore.